\documentclass[aps,twocolumn,prl,showpacs,amsmath]{revtex4}
\usepackage{graphicx}

\begin{document}

\title{Unitarity-Limited Elastic Collision Rate in a Harmonically-Trapped Fermi Gas}
\author{M. E. Gehm}
\author{S. L. Hemmer}
\author{K. M. O'Hara}
\author{J. E. Thomas}
\affiliation{Physics Department, Duke University, Durham, North Carolina 27708-0305}

\date{\today}

\begin{abstract}
We derive the elastic collision rate for a harmonically-trapped
Fermi gas in  the extreme unitarity limit where the s-wave
scattering cross section is $\sigma (k) =4\pi/k^2$, with $\hbar k$
the relative momentum. The collision rate is given in the form
$\Gamma=\gamma\,I(T/T_F)$---the product of a universal collision
rate $\gamma =k_B T_F/(6 \pi \hbar)$ and  a dimensionless function
of the ratio of the temperature $T$ to the Fermi temperature
$T_F$.  We find $I$ has a peak value of $\simeq4.6$ at
$T/T_F\simeq0.4$, $I\simeq 82\,(T/T_F)^2$ for $T/T_F\leq 0.15$,
and $I\simeq 2(T_F/T)^2$ for $T/T_F>1.5$. We estimate the
collision rate for recent experiments on a strongly-interacting
degenerate Fermi gas of atoms.
\end{abstract}

\maketitle

\section{Introduction}
Recently, we~\cite{OharaScience} and several other
groups~\cite{Dieckman,Grimm,Jincondmat,Salomoncondmat} have begun
exploring the strongly-interacting regime in degenerate Fermi
gases of atoms. In these experiments, a magnetic field is applied
to the atomic samples to tune the interparticle interactions to
the vicinity of a Feshbach resonance where the scattering length
is large compared to the interparticle spacing. In this regime,
new forms of high temperature superfluidity are
predicted~\cite{Holland,Timmermans,Griffin} and strongly
anisotropic expansion has been
observed~\cite{OharaScience,Jincondmat,Salomoncondmat}. As we
pointed out in Ref.~\cite{OharaScience}, the strongly interacting
regime leads to unitarity-limited mean field interactions as well
as unitarity-limited collision dynamics. In the latter case, the
scattering cross section is the order of $4\pi/k_F^2$, where $k_F$
is the Fermi wavevector. In the unitarity limit, the collision
rate assumes a universal form and is proportional to the Fermi
energy $k_B T_F$. At sufficiently low temperatures, Pauli blocking
may suppress the unitarity-limited elastic collision rate for the
trapped gas, producing an effectively collisionless regime.
However, a theoretical study of Pauli blocking in the
unitarity-limited regime has not been presented previously, making
it difficult to accurately estimate the collision rate. The
primary purpose of this paper is to present such a treatment.

Several groups have examined the effects of Pauli blocking on the
elastic collision rate for an energy-independent cross
section~\cite{Kennedy,HollandJin,Jin}. For comparison to the
collision rates obtained with a  unitarity-limited cross section,
we begin by deriving a  formula for the collision rate in a
harmonic trap as a function of temperature for an
energy-independent cross section $\sigma$. The results reproduce
those obtained in Ref.~\cite{HollandJin} within
10\%~\cite{factor}. We then extend the treatment to include the
energy dependence of the cross section in the extreme unitarity
limit, where the zero energy scattering length $a_S$ satisfies
$|k_Fa_S|\gg 1$ and the gas is strongly interacting. We show that
the numerically calculated collision rates for both the
energy-independent and unitarity-limited cross-sections agree with
analytic expressions derived for the high-temperature limit.

The final part of the paper defines a hydrodynamic parameter $\phi
=\Gamma/\omega_\perp$, the ratio of unitarity-limited collision
rate $\Gamma$ to the transverse oscillation frequency
$\omega_\perp$ of atoms in the trap.  We  calibrate $\phi$ by
observing the threshold for hydrodynamic expansion of a
strongly-interacting Fermi gas as a function of evaporation
time~\cite{OharaScience}. We estimate $\phi$ for several recent
experiments on strongly interacting Fermi
gases~\cite{OharaScience,Jincondmat,Salomoncondmat}.

\section{Calculating the Collision Rate}
We consider a simple model that assumes s-wave scattering is
dominant~\cite{Williams}. In this case, the collision cross
section takes the form
\begin{equation}
\sigma(k)=\frac{4\pi a_S^2}{1+k^2a_S^2},
\label{eq:crossec}
\end{equation}
where $k$ is the relative wavevector of a colliding pair of spin
up and spin down fermionic atoms.

In the trap, the average collision rate per particle, $\Gamma$, is
determined from the s-wave Boltzmann equation~\cite{Walraven}
under the assumption of sufficient ergodicity. We consider the
rate for the process in which a spin up and a spin down atom of
total energy $\epsilon_{\text{in}}=\epsilon_3 +\epsilon_4$ collide
to produce atoms with total energy
$\epsilon_{\text{out}}=\epsilon_1 +\epsilon_2$. The effects of
Pauli blocking are included for the particles on the outgoing
channel, and we assume a 50-50 mixture of atoms in the two spin
states. The depletion term in the Boltzmann equation for the
particle of energy $\epsilon_4$  is integrated over $\epsilon_4$
to determine the collision rate $\Gamma$ for either spin state (as
a collision inherently includes one atom of each spin). For an
energy-independent cross-section, the integrated loss rate is then
$\dot{N}/2\equiv -\Gamma N/2$, and
\begin{eqnarray}
\label{eq:intlossrate}
\Gamma \frac{N}{2}&=&\frac{M \sigma}{\pi^2 \hbar^3}\int d \epsilon_1 d \epsilon_2 d \epsilon_3 d \epsilon_4 \,\mathcal{D}(\epsilon_{\text{min}})\;\times \\
& &\delta (\epsilon_1+\epsilon_2-\epsilon_3-\epsilon_4) (1-f_1) (1-f_2) f_3 f_4\nonumber,
\end{eqnarray}
where $\Gamma$ is the number of collisions per second per atom and
$N$ is the total number of atoms in the trap. Here,
$\mathcal{D}(\epsilon_{\text{min}})$ is the density of states
evaluated at the energy
$\epsilon_{\text{min}}=\text{min}\{\epsilon_1,\epsilon_2,\epsilon_3,\epsilon_4\}$.
$f_i=1/(g_i+1)$ is the occupation number with $g_i=\exp
[(\epsilon_i-\mu)/k_BT]$, and $\mu$ the chemical
potential~\cite{Butts}. At zero temperature, the chemical
potential is given by the Fermi energy
$\mu(0)=\epsilon_F=(3N)^{1/3}\hbar\bar{\omega}=k_BT_F$, where
$\bar{\omega}=(\omega_\perp^2\omega_z)^{1/3}$ with $\omega_\perp$
and $\omega_z$ the transverse and axial trap oscillation
frequencies of a cylindrically symmetric trap.

For fermions, the collision cross section of Eq.~\ref{eq:crossec}
for $|ka_S|\ll 1$ is half that for bosons, i.e., $\sigma=4\pi
a_S^2$. We begin by determining $\Gamma$ for this case.
\subsection{Energy-Independent Cross-Section}
The integrand in Eq.~\ref{eq:intlossrate} is readily shown to be
symmetric under the interchange of all four particle labels.
Hence, without loss of generality, we  multiply the integrand by
4, and take $\epsilon_{\text{min}}=\epsilon_1$, and ${\cal
D}(\epsilon_{\text{min}})=[\epsilon_1^2/(2\hbar^3\bar{\omega}^3)]\theta_{21}\theta_{31}\theta_{41}$,
where $\theta_{21}=\theta (\epsilon_2-\epsilon_1)$ is a unit step
function.

It is useful to write the collision rate as the product of a
natural collision rate $\gamma_{\text{EI}}$, which depends on the
trap parameters, and a dimensionless integral
$I_{\text{EI}}(T/T_F)$, which describes the
temperature-dependence,
\begin{equation}
\Gamma=\gamma_{\text{EI}}\,I_{\text{EI}}(T/T_F)
\label{eq:collrate_constxsec}
\end{equation}

We take the natural collision rate to be the classical collision rate at $T=T_F$,
\begin{equation}
\gamma_{\text{EI}}=\frac{NM\bar{\omega}^3\sigma}{4\pi^2k_BT_F}.
\label{eq:gammaEI}
\end{equation}
Note that the rate is 1/4 of that obtained in a spin-polarized Bose gas.
With this choice, $I_{\text{EI}}$ becomes
\begin{multline}
\label{eq:Pauli}
I_{\text{EI}}(T/T_F)=144\,\iiint_0^\infty dx_1\,dx_2\,dx_3\,x_1^2\,f(x_1+x_2)\,\times \\
 f(x_1+x_3)\,[1-f(x_1)]\,[1-f(x_1+x_2+x_3)].
\end{multline}
Here $f(x)=1/[g(x)+1]$, where
$g(x)=\exp[(T_F/T)(x-\mu/\epsilon_F)]$. We assume that for the
cases of interest, the trap depth is large compared to
$\epsilon_F$ and $k_BT$.

\begin{figure}
\begin{center}
\includegraphics[width=3.0in]{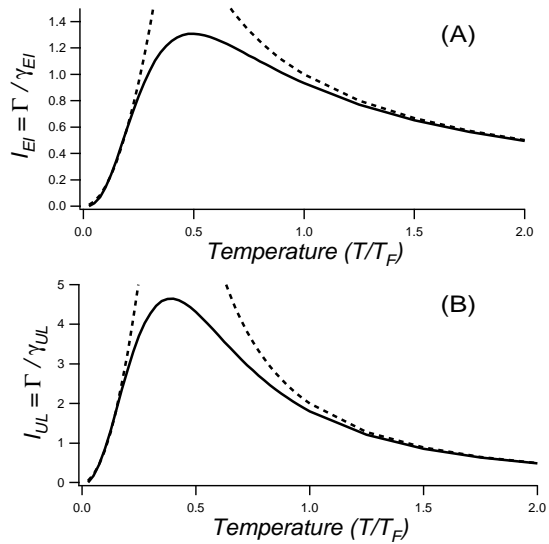}
\end{center}
\caption{\label{fig:1}Temperature dependence of the elastic
scattering collision rate $\Gamma$ in units of the natural
collision rate $\gamma$. The dashed lines indicate the high- and
low-temperature approximations. A) The collision rate for an
energy-independent cross-section. B) The collision rate for a
unitarity-limited cross-section.}
\end{figure}

$I_{\text{EI}}$ is readily determined by numerical integration
using standard results for the chemical potential as a function of
$T/T_F$~\cite{Butts}. At low temperature, $T/T_F\leq 0.2$, we find
that $I_{\text{EI}}$ is well fit by $I_{\text{EI}}(T/T_F)\simeq
15\,(T/T_F)^2$, which displays the  quadratic dependence  expected
for Pauli blocking in both final states. At high temperature,
$T/T_F>1.5$, we find the expected temperature-dependence,
$I_{\text{EI}}(T/T_F)\simeq T_F/T$ as shown below. The complete
function $I_{\text{EI}}(T/T_F)$ is plotted in Fig.~\ref{fig:1}A.
The maximum value, $I_{\text{EI}}\simeq1.3$ occurs for
$T/T_F\simeq0.5$. This demonstrates that $\gamma_{\text{EI}}$ is
essentially the maximum collision rate.

\subsection{Unitarity-Limited Cross-Section}
To include the energy dependence of the cross-section, we adopt
the notation of Ref.~\cite{Walraven}, and make the replacement
\begin{equation}
\sigma\mathcal{D}(\epsilon_{\text{min}})\rightarrow\frac{2\pi
M}{(2\pi\hbar)^3}\int_{U(\mathbf{x})< \epsilon_{\text{min}}}
d\mathbf{x} \int_{P_-(\mathbf{x})}^{P_+(\mathbf{x})}dP
\,\sigma(q), \label{eq:unitarityReplacement}
\end{equation}
where $2q=\sqrt{P_+^2  + P_-^2 -P^2}$ determines  the relative
wavevector $q$,
$P_\pm=\sqrt{2M[\epsilon-\epsilon_{\text{min}}-U(\mathbf{x})]}\pm
\sqrt{2M[\epsilon_{\text{min}}-U(\mathbf{x})]}$, and $\epsilon$ is
the total energy of the colliding particles.

We are interested in the extreme unitarity limit, where $\sigma
(k)=4\pi/k^2$ according to Eq.~\ref{eq:crossec} and the elastic
collision rate is the maximum possible. We  write the collision
rate as
\begin{equation}
\Gamma=\gamma_{\text{UL}}\,I_{\text{UL}}(T/T_F). \label{eq:collrate_unitarity}
\end{equation}
The natural collision rate, Eq.~\ref{eq:gammaEI} with $\sigma
=4\pi/k_F^2$,  where $k_F^2=2Mk_BT_F/\hbar^2$, then takes the form
\begin{equation}
\gamma_{\text{UL}}=\frac{\epsilon_F}{6\pi\hbar}. \label{eq:natgammaunitarity}
\end{equation}

The dimensionless integral $I_{\text{UL}}$ is similar to that of
Eq.~\ref{eq:Pauli},
\begin{multline}
\label{eq:IUL1}
I_{\text{UL}}(T/T_F)=144\,\int_0^\infty dx_1\,\int_0^\infty dx_2\,\int_0^\infty dx_3\,x_1^2\,\times \\
f(x_1+x_2)\,f(x_1+x_3)\,[1-f(x_1)]\,\times \\
[1-f(x_1+x_2+x_3)]\,F(2x_1+x_2+x_3,x_1),
\end{multline}
where $F(x,x_m)$ determines the energy-dependent cross section
$\sigma (q)=4\pi /q^2$ in units of $4\pi/k_F^2$. The arguments of
$F(x,x_m)$ are $x=\epsilon/\epsilon_F$, and
$x_m=\epsilon_{\text{min}}/\epsilon_F$. As in Eq.~\ref{eq:Pauli},
we take $x_m=x_1$ without loss of generality.

For a harmonic potential,
\begin{equation*}
F(x,x_m)=\frac{16}{\pi\sqrt{2x_m}}\int_0^1\frac{du\,u^2}{\sqrt{x-2x_mu^2}}\,
\ln\left[\frac{\chi_{+}\left(\alpha,u\right)}{\chi_{-}\left(\alpha,u\right)}\right],
\label{eq:Ffunc}
\end{equation*}
where $\chi_{\pm}(\alpha,u)=A(\alpha,u)\pm B(\alpha,u)$,
$A(\alpha,u)=\alpha+2(1-2u^2)$,
$B(\alpha,u)=2\sqrt{2(1-u^2)(\alpha-2u^2)}$, and $\alpha=x/x_m$.

We can simplify the form of the integrals in Eq.~\ref{eq:IUL1} by
transforming from coordinates $\{x_1,x_2,x_3,u\}$ into coordinates
$\{w,y,z,s\}$, where $w=x_1$, $y=(x_2+x_3)/2x_1$, $z=(x_3-x_2)/2
x_1$, and $s=1-u^2$. After the transformation we have
\begin{multline}
I_{\text{UL}}(T/T_F)=144\,\int_0^\infty dy\, G(y)\int_0^\infty dw\,w^3\,[1-f(w)]\,\times\\
\left[1-f[w(1+2y)]\right]\int_{-y}^{y}dz \,f[w(1+y+z)]\,\times\\f[w(1+y-z)]
\label{eq:IUL2}
\end{multline}
with $G(y)$ given by
\begin{multline}
G(y)=\frac{16}{\pi}\int_0^1 ds \,\sqrt{\frac{1-s}{s+y}}\,\times\\
[\ln (s+y/2+\sqrt{s(s+y)})-\ln (y/2)].
\label{eq:Gfunc}
\end{multline}

Eq.~\ref{eq:IUL2} is plotted in Fig.~\ref{fig:1}B. The maximum
collision rate is larger by a factor $\simeq 4$ than for an energy
independent cross section $\sigma =4\pi/k_F^2$, consistent with
estimates from radio-frequency  measurements of the mean field
shift in Ref.~\cite{Jincondmat}. For $T/T_F<0.15$, we again find
the quadratic temperature-dependence that results from Pauli
blocking, $I_{\text{UL}}\simeq 82\,(T/T_F)^2$.   In the
high-$T/T_F$ limit we find $I_{\text{UL}}\simeq 2\,(T_F/T)^{2}$.
This matches the high-temperature prediction given below. The
maximum value of $I_{\text{UL}}\simeq4.6$ occurs at $T/T_F\simeq
0.4$.

\subsection{Comparison With Analytic High-$T/T_F$ Results}
We check the numerical results for the temperature dependence of
the rates by calculating the collision rate $\Gamma_{\text{HT}}$
for $T\gg T_F$ directly from the phase-space s-wave Boltzmann
equation~\cite{Walraven} in the high-$T/T_F$ limit, where the
occupation number is given by a Boltzmann factor and
Pauli-blocking can be neglected. Including the dependence of the
scattering cross section on the relative speed $v_r$, we find
generally that
\begin{equation}
\Gamma_{\text{HT}}\, \frac{N}{2}=\int
d{\mathbf{x}}\,n_{\uparrow}(\mathbf{x})n_{\downarrow}(\mathbf{x})\langle v_r\sigma (v_r)\rangle ,
\label{eq:classical}
\end{equation}
where the angled brackets denote an average over the relative
velocity distribution for pairs of atoms and $\int
d{\mathbf{x}}\,n_{\uparrow ,\downarrow}(\mathbf{x})=N/2$.
Eq.~\ref{eq:classical} has a simple physical interpretation: The
spin up atoms at position $\mathbf{x}$, i.e.,
$n_{\uparrow}(\mathbf{x})\,d{\mathbf{x}}$, are hit by spin down
atoms at a rate $n_{\downarrow}(\mathbf{x})\langle v_r\sigma
(v_r)\rangle$. For an energy-independent cross section, we obtain
$\Gamma_{\text{HT}} =\gamma_{\text{EI}}\,T_F/T$ in agreement with
the numerical results for $I_{\text{EI}}(T/T_F)$ of
Fig.~\ref{fig:1}A. The temperature dependence in this limit arises
from the flux, which is the product of the density and the
relative velocity, i.e., $n\langle v_r\rangle\propto 1/T$. For the
extreme unitarity-limited cross section $\sigma=4\pi/k^2$, with
$\mu v_r=\hbar k$, we obtain $\Gamma_{\text{HT}}
=\gamma_{\text{UL}}\,2(T_F/T)^2$ in agreement with the numerical
results for $I_{\text{UL}}(T/T_F)$ shown in Fig.~\ref{fig:1}B. In
this case, the $1/T^2$ temperature dependence arises because both
the flux and  the cross section vary as $1/T$.
\section{Hydrodynamic Parameter $\phi$}

The collisional state of a gas can be described by a hydrodynamic
parameter $\phi$ which is the number of collisions that an atom
experiences during a characteristic timescale. When $\phi$ is
large, the gas is collisionally hydrodynamic, while when $\phi$ is
small, the gas is collisionless. The actual values of $\phi$ which
qualify as ``large'' or ``small'' are determined by calibration.

We choose the characteristic time-scale to be $1/\omega$, and take
$\phi =\Gamma/\omega$ where $\omega=\omega_\perp,\omega_z$ are the
oscillation frequencies of atoms in the trap. These are also the
natural time scales for ballistic expansion, where the size of the
cloud scales as $\sqrt{1+\omega^2 t^2}$ in each direction.  In the
unitarity-limited regime, and for a cylindrically-symmetric trap
with elongation parameter $\lambda=\omega_z/\omega_\perp$, we can
write $\phi_\perp$ as
\begin{equation}
\phi_\perp=\frac{(3\lambda N)^{1/3}}{6\pi}\,I_{\text{UL}}(T/T_F)
\label{eq:PhiDef}
\end{equation}
 with $N$ the total number of atoms in the 50-50 mixture.
 Then, $\phi_z=\phi_\perp/\lambda$.
\subsection{Calibrating $\phi$}
\begin{figure}
\begin{center}
\includegraphics[width=3.25in]{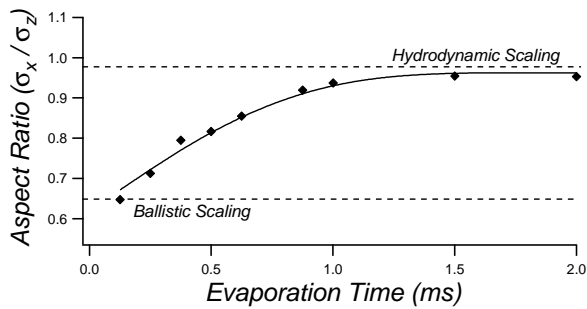}
\end{center}
\caption{\label{fig:3} Observed aspect ratio of a
strongly-interacting Fermi gas after 600 $\mu$s free expansion as
a function of evaporation time. The aspect ratios corresponding to
ballistic and hydrodynamic expansion are indicated by the dashed,
horizontal lines. The solid curve has been included to guide the
eye. The gas evolves smoothly from ballistic expansion to
hydrodynamic expansion as the evaporation time is increased.}
\end{figure}

We now turn to the question of determining the approximate value
of $\phi$ for which  the transition between collisionless and
collisional behavior occurs. In Ref.~\cite{OharaScience}, we
investigated the anisotropic expansion properties of a
strongly-interacting, degenerate Fermi gas of $^6$Li. When
released from a highly-elongated trap, the originally narrow
dimensions of the gas expanded rapidly, while the broad dimension
remained largely unchanged---inverting the aspect ratio of the
cloud. We have studied how the observed aspect ratio of the
expanded cloud varies with the duration of evaporative cooling.
The aspect ratios are measured for a fixed expansion time of 600
$\mu$s and are compared to the predictions of ballistic and
hydrodynamic expansion. The results are plotted in
Fig.~\ref{fig:3}.

From this figure, we see that the hottest clouds (short
evaporation times) expand ballistically, while the coldest clouds
(longest evaporation times) expand hydrodynamically. Ballistic
expansion is expected in a normal, collisionless gas. Since the
rapid transverse expansion extinguishes collisions before the
axial distribution can change significantly, the collisional
behavior of the expanding gas can be associated with $\phi_\perp$.
For the shortest evaporation times  shown in Fig.~\ref{fig:3}, we
observe ballistic scaling with $T=3T_F$ and $N=4\times 10^5$. For
our trap $\lambda=0.035$. From Eq.~\ref{eq:PhiDef}, we obtain
$\phi_\perp=0.4$ which then corresponds to approximately
collisionless behavior. Therefore, $\phi=0.4$ is, in general, the
condition for collisionless behavior for any timescale $1/\omega$.
\subsection{Application to Experiment}
In addition to our work in \cite{OharaScience}, several groups
have also recently observed hydrodynamic expansion of a Fermi gas
in the strongly-interacting
regime~\cite{Jincondmat,Salomoncondmat}. The authors of these
papers claim that their experiments are collisionally
hydrodynamic. For Ref.~\cite{Jincondmat}, we take the total number
of atoms $N=2.4\times 10^5$ and $T/T_F=0.34$, with $\lambda
=0.016$, yielding $I_{UL}=4.5$ and $\phi_\perp =5.4$. For
Ref.~\cite{Salomoncondmat}, we take $N=7\times 10^4$, $T/T_F=0.6$
and $\lambda =0.35$, yielding $I_{UL}=3.5$ and $\phi_\perp =7.7$.
Hence, we agree with their conclusions.

Our experiments~\cite{OharaScience} produce strongly-interacting
Fermi gases at considerably lower temperatures. In those
experiments, the total number of atoms is $N\simeq 1.5\times10^5$
for temperatures $0.08\leq T/T_F \leq 0.2$.  Eq.~\ref{eq:PhiDef}
yields $0.7\leq \phi_\perp\leq 3.7$, while $20 \leq \phi_z \leq
106$ . The values of $\phi$ corresponding to our lowest
temperatures indicate that the trapped gas is nearly collisionless
on the transverse timescale, but collisional on the axial.

The onset of high-temperature superfluidity has been recently
predicted in the temperature range
$T/T_F=0.25-0.5$~\cite{Holland,Timmermans,Griffin}. Since Pauli
blocking is ineffective for the unitarity limited cross section
when $T/T_F\geq 0.25$, it is not clear how collisions in the
normal component will affect the formation of this high
temperature superfluid.

For an  expanding gas, we cannot make a definitive statement about
the collisional nature, even if it were collisionless when
trapped, as Pauli blocking may become ineffective as a result of
nonadiabaticity in the expansion, deformation of the Fermi
surface, or through other effects~\cite{KetterleHoPrivate}. The
magnitude of these effects, and to what extent they modify $\phi$
remains an open question. We are therefore working on experiments
which will directly determine if the gas contains a superfluid
fraction.

We are indebted to J. T. M. Walraven for stimulating
correspondence regarding this work. This research is supported by
the Physics Divisions of the Army Research Office and the National
Science Foundation, the Fundamental Physics in Microgravity
Research program of the National Aeronautics and Space
Administration, and the Chemical Sciences, Geosciences and
Biosciences Division of the Office of Basic Energy Sciences,
Office of Science, U. S. Department of Energy.


\begin{thebibliography}{99}
\bibitem{factor}We take into account a factor of two difference between our definition of the collision rate and
that of Ref.~\cite{HollandJin}. We assume  the thermal energy to
be much smaller than the trap depth while Ref.~\cite{HollandJin}
assumes a truncated distribution to treat evaporation.
\end{thebibliography}
\end{document}